\documentclass[prl,twocolumn,aps,superscriptaddress,showpacs,floatfix]{revtex4}
\usepackage{amsmath}
\usepackage{graphicx}

\begin{document}
\title{Noise limited computational speed}
\author{L. Gammaitoni}
\affiliation{Dipartimento di Fisica, Universit\`a di Perugia,
I-06123 Perugia, Italy, and Istituto Nazionale di Fisica Nucleare,
Sezione di Perugia, I-06123 Perugia, Italy}
\date{\today}

\begin{abstract}
In modern transistor based logic gates, the impact of noise on computation has become increasingly relevant since the voltage scaling strategy, aimed at decreasing the dissipated power, has increased the probability of error due to the reduced switching threshold voltages. In this paper we discuss the role of noise in a two state model that mimic the dynamics of standard logic gates and show that the presence of the noise sets a fundamental limit to the computing speed. An optimal {\it idle time} interval that minimizes the error probability, is derived.

\end{abstract}
 \maketitle

The role of noise in computation devices has become increasingly relevant both in the quantum\cite{quantum1,quantum2} and in the classical\cite{Sano, Kish} regime.
With the present tendency to scale down CMOS based devices toward the nano-meter region\cite{Hanson,[3]}, the noise immunity in a low energy dissipation scenario has become the recurring objective of significant research efforts in this field \cite{[5],[6]}. Some authors have focused their attention on the potential role of noise in nanoscale devices where noise driven dynamics \cite{[7]} has been invoked to explain the experiments and to optimize future design \cite{ [8]}. In order to address a non-negligible error probability a number of strategies have been devised where a probabilistic approach to the computational task has been often invoked \cite{ [9]}. In this letter we focus our attention on the very basic mechanisms of the switch dynamics that are responsible of the functioning of traditional transistor based logic gates, with the aim of clarifying the impact of noise on computation errors.

Noise can affect the functioning of computing devices in a number of different ways. To fix our ideas let's consider a simple logic gates that constitute the building block of complex networks aimed at realizing computing tasks in modern electronic devices. Here the noise has two deleterious effects: first, it can interact with an unperturbed static signal causing the loss of information carried by the static node of the computational network; second, it can affect the functioning of a switching node by altering its dynamical properties (e.g.: slew, delay). In this letter we deal with the second effect. 
More precisely, we focus our attention on the very basic mechanism of the switching event in a logic gate. Reduced to the essential this mechanism can be sketched as an output change in response to a threshold-crossing event. For the sake of simplicity we consider here the simpler of the various switching computing elements: the Logic Inverter or NOT gate. This gate is usually operated as a pure switching device, governed by the following rule: the output logic state commutes from $1$ (or HIGH) to $0$ (or LOW) if the input signal crosses from below the upper switching threshold $b_u$ (transition from LOW to HIGH). As shown in Fig.1 (left hand side) a time delay between input and output occurs: before the output signal is stable in the desired logic state, some time is required after the application of the input signal. The amount of such a delay, called {\it propagation delay}, $t_p$, characterizes the different {\it Logic Families} (TTL, CMOS, ECL,...) and ranges between few ns and few tens of ns. A significant contribution to the propagation delay is given by the rise time $t_r$ that in turn affects what is usually called the {\it slew rate} of the device. The separation voltage between the up and down thresholds, $b_u - b_d$, depends on the different {\it Families} and ranges from $0.7$ V in {\it ECL logic} to around $28$ V in {\it relay logic}.

\begin{figure*}[ht]
\includegraphics*[width=19.0cm]{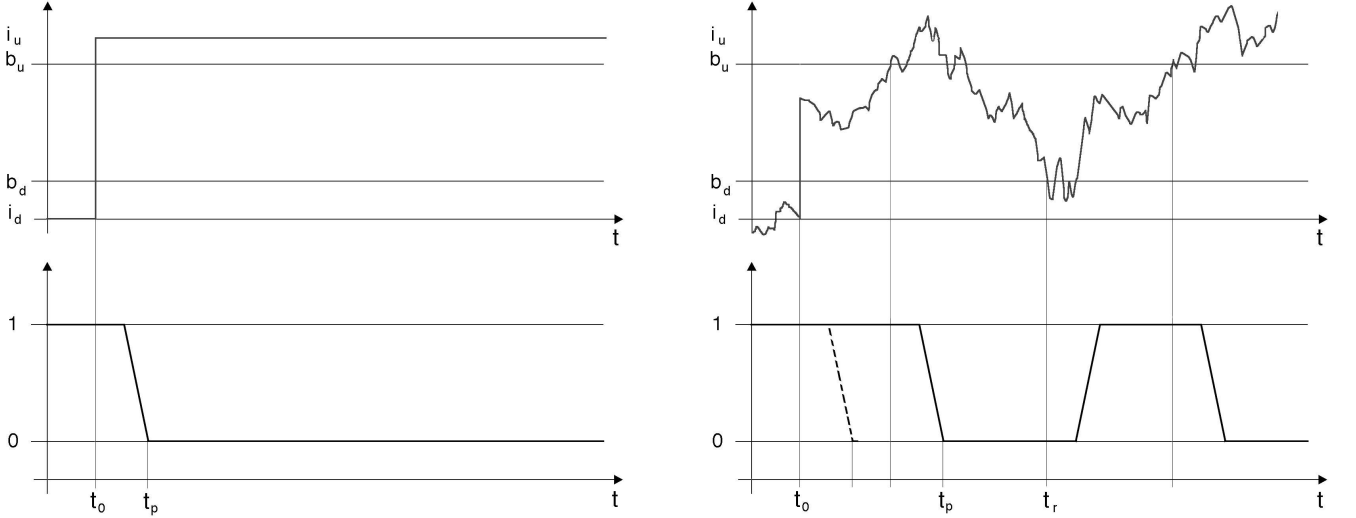}
\caption{\label{F1} 
Time diagram of a Logic Inverter (NOT gate). Left hand side: input (upper trace) and output (lower trace) time series for the case where there is no noise present at the input.  Right hand side: input (upper trace) and output (lower trace) time series for the case where the input signal is affected by additive noise. The output time series shows two typical phenomena: 1) a delayed switching due to the trapping effect\cite{trapping} that introduces a wait time that adds to $t_p$ 2) at later time $t_r$ a re-crossing phenomena resulting in a bit-flip error.
}
\end{figure*}

A number of different noise induced phenomena, ranging from switching delays (see e.g. {\it noise on timing} and {\it noise-on-delay} effects) to {\it bit-flip} errors, threaten the correct functioning of threshold-crossing based logic gates. Main physical noise sources being power supply noise, environmental noise and also thermal noise when the devices dimensions hit the nanoscale.
In order to model the dynamical effects of the noise on the switching mechanism we sketched in fig.1 (right hand side) a common scenario. Here, the time diagram shows the input and output time series for the case where the input signal is affected by noise of intensity comparable with the threshold separation. For generality purpose we considered the case of exponentially correlated, Gaussian distributed, stationary noise with correlation time $\tau$ and standard deviation $\sigma$. This noise is added to the deterministic signal shown in the leftmost part of the figure and the resulting signal is presented in the upper diagram. The effect of the noise in the gate response (output time series, lower diagram) is twofold: $1)$ it can initially prevent the input signal from crossing the relevant threshold ($b_u$ in the example) postponing in time this event and thus resulting in a longer propagation delay $t_p$ ({\it delayed switching error}). $2)$ Once the device switching is completed, it might cause a re-crossing of the opposite threshold ($b_d$ in the example) causing a {\it bit-flip error}.

In the following we will analyze in detail the statistics of these two events that directly reflects into the computational error probability.

{\bf 1)} {\it delayed switching}. The {\it delayed switching error} is produced when the NOT gate, expected to be in the LOW state is found instead still in the HIGH state due to a delayed switch. This error is clearly time dependent and we are interested in estimating how its probability evolves with time. In order to have a switch delayed, two conditions have to be met: 

$a)$ at time $t=t_0$ , due to the presence of noise, the input signal of amplitude $i_u$ that makes the device commute from HIGH to LOW cannot reach the switching threshold $b_u$. This happens when $i_u + \xi_0 <b_u$, or $\xi_0 <b_u - i_u = b_e$, where $\xi_0 = \xi(t_0)$ is the instantaneous value of the noise (a realization of the stochastic process $\xi(t)$ sampled at $t=t_0$). Such an event happens with probability:

\begin{equation}
\label{P1a} P_{1a} = \Phi\left(\frac {b_e} {\sigma}\right) =  \frac{1} {2} (1+ \mbox{Erf}(\overline{b_e})
\end{equation}

where $\mbox{Erf}(x)$ indicates the {\it Error-function} and $\overline{b_e} = b_e/( \sqrt{2\sigma^2})$. 

$b)$ At time $t > t_0$ the noise is such that the condition $\xi(t) <b_u - i_u$ still holds. This second condition is satisfied with probability $P_{1b}$ that can be estimated as follows\cite{PRE}. 
Once the condition $a)$ is satisfied $(\xi_0<b_e)$ it can take some time before the input signal reaches the upper threshold $b_u$. This delay can be estimated by considering the so-called 
{\it First Passage Time} (FPT), i.e. the time the stochastic process $\xi(t)$ takes to reach $b_e$ (i.e. to go from $\xi(t_0) <b_e$ to $b_e$ with absorbing boundary in $b_e$ and reflecting 
boundary in $- \infty$). This delay is a random variable $t$ whose mean value $<t>=T_1$ is called MFPT and whose probability density function $p_1(t)$ is exponential\cite{Fokker, Marchesoni}. 
The error probability $P_{1b}$ coincides with the probability that in the time interval $[t_0 , t]$ there was no crossing of $b_e$, i.e.:

\begin{equation}
\label{P_1b} 
P_{1b}(t_0,t) \equiv 1- \int^{t}_{t_0} p_1(t) \,dt  = e^{-\frac{(t - t_0)}{T_1}}
\end{equation}

The relevant time $T_1$ is a function of the noise characteristics\cite{PRE}:
\begin{equation}
\label{T_1}
T_1 (\overline{b_{e}}) = \frac{\tau}{N} \int_{-\infty}^{\overline{b_{e}}}\int_{z}^{\overline{b_{e}}}e^{-z^2+x^2} (1+\mbox{Erf}(x))\;dx\;dz 
\end{equation}

Where $N =  \frac{1} {2} (1- \mbox{Erf}(\overline{b_{e}}))$.

Finally the {\it delayed switching error} probability is obtained by the combination of the two error probabilities: 

\begin{equation}
\label{P_1}
P_1 (t_0 , t) = P_{1a} P_{1b} = \Phi\left(b_e/\sigma\right) e^{-\frac{(t - t_0)}{T_1}}. 
\end{equation}

Having obtained the expression for the error probability $P_1$ we can now derive a useful prediction for operating the NOT gate in noisy conditions. In Fig.2 the {\it delayed switching error} probability $P_1$ is shown as a function of $t /\tau$. As expected this probability decreases with time and becomes negligible in the long time. If we fix what we consider an acceptable error probability $\epsilon$, than we can easily compute a safe {\it wait time} $t_w$ after which the error probability stays below $\epsilon$, i.e. $P_1 <\epsilon$ when  $t > t_w$. The relation between $t_w$ and $\epsilon$ is easily obtained from eq. (\ref{P_1}) as

\begin{equation}
\label{t_w}
t_w = T_1\; ln\left(\frac{\Phi(b_e/\sigma)}{\epsilon}\right)
\end{equation}

where for simplicity we have assumed $t_0 = 0$. Most notably, if we are willing to accept an error probability $\epsilon = \Phi(b_e/\sigma)$   or greater, the wait time $t_w$ amounts to zero.

\begin{figure}[ht]
\includegraphics*[width=8.5cm]{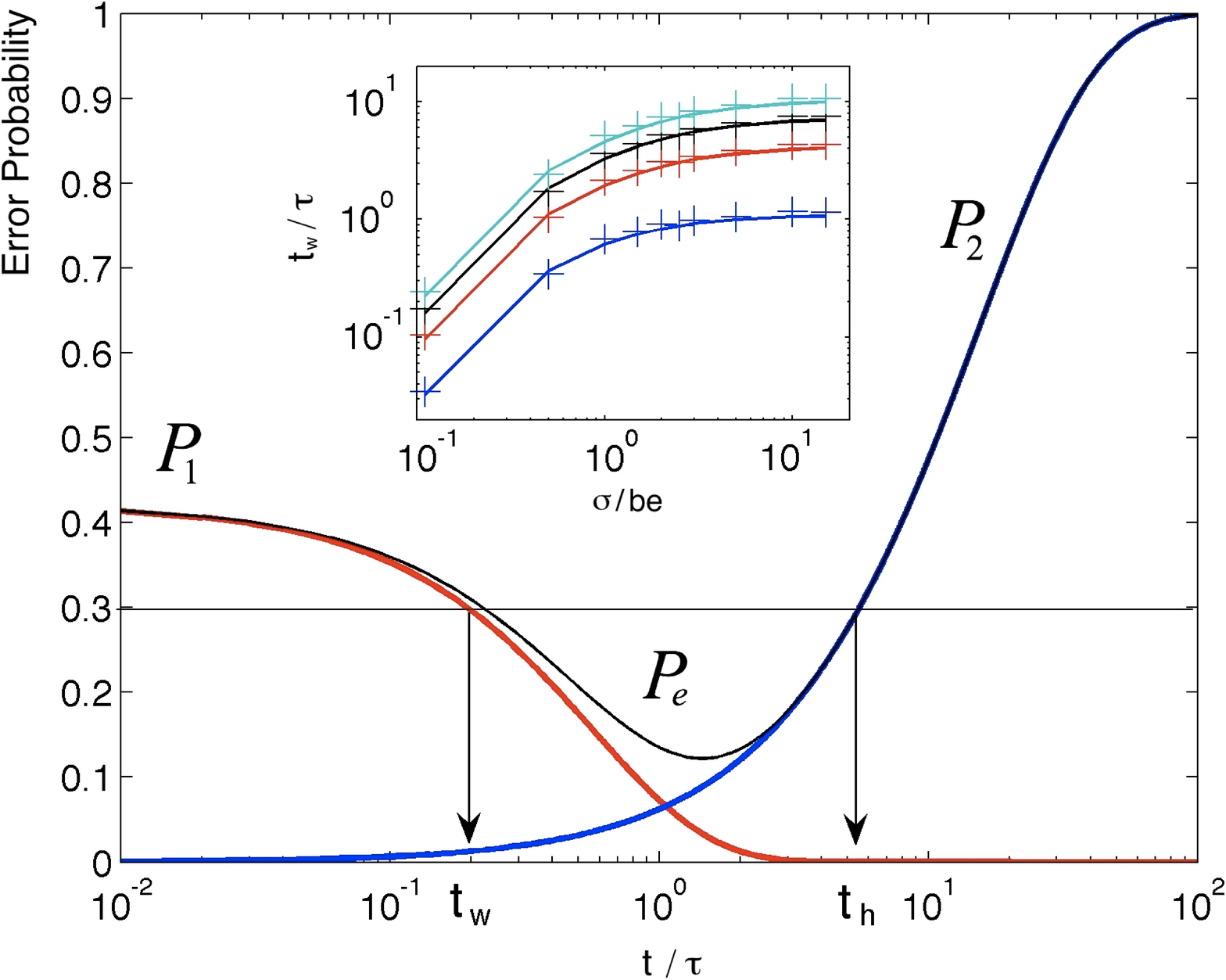}
\caption{\label{F1} (Color online) 
Computational error probability. The {\it delayed switching error} probability $P_1$ is shown as a function of $t/\tau$ (red online) together with the bit-flip error probability $P_2$ (blue online) and the resulting total error probability $P_e$ (black online). Parameter values: $\tau = 10^{-9}$ s and $\sigma = 1$ V, $i_u = 4.2$ V, $b_u = 4.0$ V, $b_e = 0.2$ V. As an example, an error probability $\epsilon = 0.3$  line is drawn across the curves. The intercepts at $t_w$ and $t_h$ respectively are drawn (down arrows). Inset: normalized wait time $t_w/\tau$ versus $\sigma/b_e$ for different values of the error probability (from above): $\epsilon = 10^{-1}$ (green), $\epsilon = 10^{-3}$ (black), $\epsilon = 10^{-5}$ (red), $\epsilon = 10{-7}$ (blue). Theoretical predictions (continuous line) are in close agreement with digital simulation (crosses).
}
\end{figure}

{\bf 2)} {\it bit-flip}: Operationally, notwithstanding the {\it delayed switching error}, it would seem that we can still use the NOT gate with a negligible error probability, provided we are willing to wait long enough (longer than $t_w$). Unfortunately there is another error that comes into play if we wait too long: the {\it bit-flip error}. As shown in Fig.1, after a switch event (HIGH to LOW) is occurred, a new unwanted switch can occur in the opposite direction (LOW to HIGH), if the noise assumes a value $\xi(t) <b_d - i_u  = c_e$ at a time $t$, while the input signal is still $i_u$ . For practical purposes also a {\it bit-flip error} of short duration is deleterious to the signal integrity and can seriously compromise the functioning of the logic gate. To estimate the {\it bit-flip error} probability $P_2$, let's assume that at $t=t_0$ there is a switch event (HIGH to LOW), i.e.: $\xi(t_0) \ge b_e$ . We are interested in computing the time $t$ the stochastic process $\xi(t)$ takes to reach $c_e$ (i.e. to go from $\xi(t_0) \ge b_e$ to $c_e$ with absorbing boundary in $c_e$ and reflecting boundary in $+ \infty$). This time $t$ is a random variable whose mean value is $T_2$ (MFPT) and whose probability density function $p_2(t)$ is exponential\cite{PRE}. For what we said, $P_2(t_0 , t)$ represents the probability that there was a crossing of $c_e$ in the time interval $[t_0 , t]$. 

\begin{equation}
\label{P_2} 
P_2(t_0 , t) \equiv \int^{t}_{t_0} p_2(t) \,dt = 1 - e^{-\frac{t - t0}{T_2}}.
\end{equation}

 The relevant time $T_2$ can be computed as\cite{PRE}:
 
\begin{equation}
\label{T_2}
T_2 (\overline{b_{e}}, \overline{c_{e}}) = \frac{\tau}{N} \int^{+\infty}_{\overline{b_{e}}}\int^{z}_{\overline{c_{e}}}e^{-z^2+x^2} (1-\mbox{Erf}(x))\;dx\;dz 
\end{equation} 

In Fig.2 $P_2$ is shown as a function of $t /\tau$. As expected this probability increases with time and approaches unity when $t$ grows to infinity. 
  
For the {\it bit-flip error}, once we fix an acceptable error probability $\epsilon$, we obtain a safe {\it hurry time} $t_h$ before which the error probability stays below $\epsilon$. The relation between $t_h$ and $\epsilon$ is easily obtained from eq. (\ref{P_2}) as

\begin{equation}
\label{t_h}
t_h = - T_2\; ln(1-\epsilon), 
\end{equation}

where we have assumed $t_0 = 0$.

Finally, if we take into account the two errors previously discussed, we are now in position to express the total error probability: $P_e = P_1 + P_2$ . $P_e$ is also shown in Fig.2. It is apparent that $P_e$ has a minimum for $t=t_m$ with $t_w <t_m <t_h$. Operatively, if we fix an acceptable error probability $\epsilon$ this identifies an idle time interval $(t_{is}, t_{ie})$ of amplitude $\Delta T_i = t_{ie} - t_{is}$, where the total error probability $P_e$ is smaller than $\epsilon$. When $T_1 \ll T_2$ we can approximate $t_{ie}$ with $t_h$ and $t_{is}$ with $t_w$, thus $\Delta T_i  \simeq t_h - t_w$. 
It is worth noticing that one of the consequences of this analysis is that $P_e$ assumes a minimum value identified by the condition $t_{is} = t_{ie} = t_m$. This implies that when operating a logic gates in the presence of noise, the probability of error cannot be made arbitrarily small but only as small as $\epsilon_m = P_e(t_m)$. Remarkably $\epsilon_m $ does not depend on the noise correlation time but only on the noise intensity\cite{gammaitoni}.

\begin{figure}[ht]
\includegraphics*[width=8.5cm]{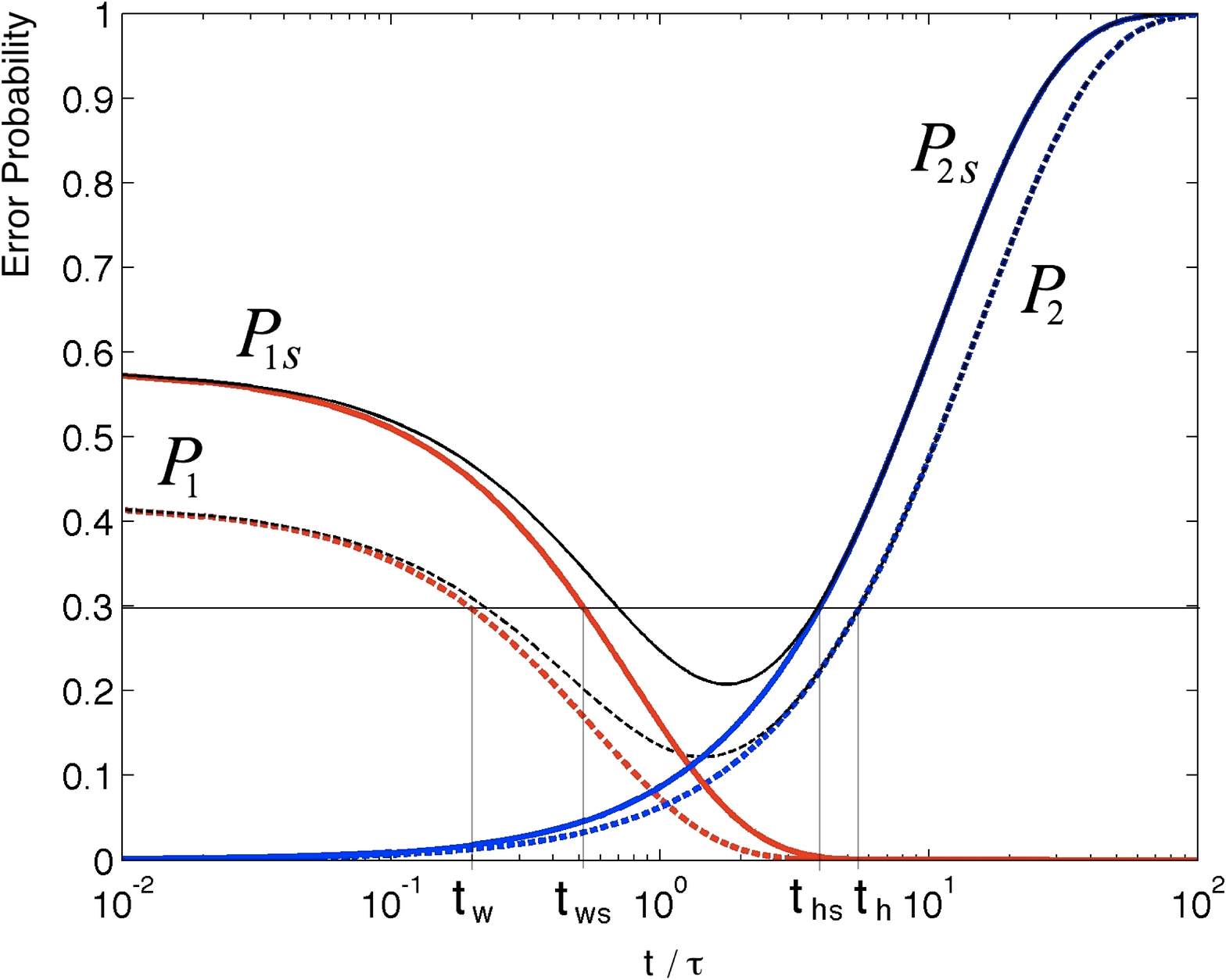}
\caption{\label{F3} (Color online) 
Computational error probability for the supra-threshold case: comparison with sub-threshold driving. The {\it delayed switching error} probabilities $P_1$ (dashed) and $P_{1s}$ (continuous) are shown as a function of $t/\tau$ together with the {\it bit-flip error probabilities} $P_2$ and $P_{2s}$ and the resulting total error probabilities $P_e$ and $P_{es}$. For comparison with Fig. 2 an error probability $\epsilon = 0.3$  line is drawn across the curves. The intercepts at $t_w$, $t_{ws}$, $t_h$ and $t_{hs}$ respectively are drawn (down arrows). 
}
\end{figure}

The role of noise in computing devices however can also be seen from a different perspective. Instead of being a mere disturbance it can be considered as an essential part in the computing process itself. This is the case for example, when we consider sub-threshold gate driving, i.e. when $i_u <b_u$. In the absence of noise no switch is possible and the gate cannot operate. Instead, also a noise of small intensity can bring (in due time) the input signal above the threshold and thus drive the gate for the computing task. Scenarios where the noise can play a beneficial role are not new in the literature; see e.g. the {\it Stochastic Resonance} phenomenon\cite{SR} or the {\it Dithering effect}\cite{dithering}. To compute the time evolution of the error probability $P_{1s}$ for the sub-threshold case we can proceed as we did before for the analogous quantity $P_1$. We obtain: 

\begin{equation}
\label{P_1s}
P_{1s}(t) = P_{1as} P_{1bs} = \Phi_{be}\; e^{-\frac{t}{T_{1s}}}. 
\end{equation}

The main difference being that in this case $b_u - i_{us} = b_e  > 0$. Moreover, while $T_1$ is a monotonic growing function of $\sigma$, $T_{1s}$, the MFPT for this process, is a monotonic decreasing function of $\sigma$ and $T_{1s} > T_1$ for any value of $\sigma$\cite{gammaitoni}.
The derivation of the {\it bit-flip error} probability $P_{2s}$ is made according to the derivation of $P_{2}$ for the supra-threshold case. In Fig.3 the error probabilities for the two scenarios (supra- and sub-threshold) are compared. Noticeably, for a given acceptable error probability the following relation holds for the two corresponding idle interval: $t_w <t_{ws }<t_{hs} <t_h$. However, in the large noise intensity limit ($\sigma \gg |b_e|$), $t_w$ and $t_{ws}$ admit the same limit and the idle time difference between the two cases becomes negligible.

In conclusion we have shown that the presence of noise of intensity comparable with the difference between the input signal amplitude and the threshold value can seriously limit the computing speed of standard logic gates. More specifically, we have demonstrated that computation in threshold based devices (e.g. transistor based logic gates) can still be performed provided that the system clock is operated accordingly to the existence of a proper idle time interval that is a function of the noise properties. Finally we have shown that in the large noise scenario, the computing device can be operated also with an energy saving sub-threshold signal. 
We anticipate this result to be potentially relevant toward the design of nano-scale computers where thermal and ambient noises, instead of being a mere source of disturbances could be useful components of the computing process.
The author gratefully acknowledge financial support from Ministero Italiano della Ricerca Scientifica (PRIN 2004) and European Commission (FPVI, STREP Contract N. 034236 SUBTLE: Sub KT Low Energy Transistors and Sensors). The author also thanks the Office of Naval Research for support during the initial phase of this research.

\end{document}